\documentclass[11pt]{article}
\usepackage{amsmath,amssymb,amsthm,amsxtra,overpic,bbm,bm,epsfig,ulem,color,multirow}
\usepackage{braket}

\begin{document}

\begin{center}
{\Large Rescuing leptogenesis in inverse seesaw models \\ with the help of non-Abelian flavor symmetries}
\end{center}

\vspace{0.05cm}

\begin{center}
{\bf Yan Shao, \bf Zhen-hua Zhao\footnote{Corresponding author: zhaozhenhua@lnnu.edu.cn}} \\
{ $^1$ Department of Physics, Liaoning Normal University, Dalian 116029, China \\
$^2$ Center for Theoretical and Experimental High Energy Physics, \\ Liaoning Normal University, Dalian 116029, China }
\end{center}

\vspace{0.2cm}

\begin{abstract}
The inverse seesaw (ISS) model provides an attractive framework that can naturally explain the smallness of neutrino masses while accommodating some sterile neutrinos potentially accessible at present or future experiments. However, in generic ISS models with hierarchical pseudo-Dirac (PD) sterile neutrino pairs, the generation of the observed baryon asymmetry of the Universe via the leptogenesis mechanism is extremely challenging. In this paper, we investigate rescuing leptogenesis in the ISS model with the help of non-Abelian flavor symmetries which have the potential to explain the observed peculiar neutrino mixing pattern: we first implement non-Abelian flavor symmetries to naturally enforce mass degeneracies among different pseudo-Dirac sterile neutrino pairs and then break them in a proper way so that resonant leptogenesis among different PD sterile neutrino pairs can arise, thus enhancing the generated baryon asymmetry. To be specific, we have considered the following two well-motivated approaches for generating the tiny mass splittings among different PD sterile neutrino pairs: one approach makes use of the renormalization-group corrections to the sterile neutrino masses, while the other approach invokes non-trivial flavor structure of the $\mu^{}_{\rm s}$ matrix. For these two scenarios, we aim to explore the viability of leptogenesis and to identify the conditions under which the observed baryon asymmetry can be successfully reproduced.
\end{abstract}

\newpage

\section{Introduction}

The Standard Model (SM) of particle physics, while being incredibly successful, fails to address two fundamental puzzles: the origin of neutrino masses and the generation of the baryon-antibaryon asymmetry of the Universe (BAU). On the one hand, neutrino oscillations have conclusively demonstrated that neutrinos have non-zero masses \cite{xing}, which contradicts the Standard Model's assumption of massless neutrinos. On the other hand, the BAU puzzle, quantified as \cite{planck}
\begin{eqnarray}
Y^{}_{\rm B} \equiv \frac{n^{}_{\rm B}-n^{}_{\rm \bar B}}{s} \simeq (8.69 \pm 0.04) \times 10^{-11}  \;,
\label{1.1}
\end{eqnarray}
where $n^{}_{\rm B}$ ($n^{}_{\rm \bar B}$) denotes the baryon (antibaryon) number density and $s$ the entropy density, remains unresolved.

One of the most popular and natural ways to generate the tiny neutrino masses is via the type-I seesaw mechanism \cite{seesaw1}-\cite{seesaw5} which introduces at least two heavy right-handed neutrinos (RHNs) $N^{}_I$ (for $I=1, 2, ...$) into the SM. With the help of this mechanism, the smallness of neutrino masses can be naturally ascribed to the heaviness of RHNs. Furthermore, the type-I seesaw model also provides an elegant solution to the BAU puzzle via the leptogenesis mechanism \cite{leptogenesis}-\cite{Lreview4}: a lepton asymmetry is first generated from the CP-violating and out-of-equilibrium decays of RHNs and then partially converted to the desired baryon asymmetry via the sphaleron processes. However, the traditional type-I seesaw mechanism and  associated leptogenesis mechanism have the drawback that the newly introduced particle states (i.e., RHNs) are too heavy to be directly accessed by foreseeable experiments: on the one hand, in order to naturally generate eV-scale neutrino masses, the RHN masses need to be around ${\cal O}(10^{13})$ GeV; on the other hand, when the RHN masses are hierarchial, in order to generate sufficient baryon asymmetry, the lightest RHN mass needs to be above ${\cal O}(10^9)$ GeV \cite{DI}.

In order to make the seesaw scale potentially accessible at present or future experiments, the inverse seesaw (ISS) model offers an elegant alternative \cite{ISS1, ISS2}. In addition to the RHNs, the model also introduces some other singlet fermions $S^{}_{I}$ (for $I=1, 2, ...$). And there are the following Lagrangian terms relevant for the generation of neutrino masses:
\begin{eqnarray}
-\mathcal{L}^{}_{\rm mass} \supset (Y^{}_{\nu})^{}_{\alpha I} \, \overline{L^{}_\alpha} \, \widetilde{H} \, N^{}_I + (M^{}_{\rm R})^{}_{IJ} \, \overline{N^c_{I}} \, S^c_{J} + \frac{1}{2} (\mu^{}_{\rm s})^{}_{IJ} \overline{S^{}_{I}} S^{c}_{J} + {\rm h.c.}
\label{1.2}
\end{eqnarray}
where $L^{}_\alpha$ (for $\alpha = e, \mu, \tau$) and $H$ (for $\widetilde{H} = {\rm i \sigma^{}_2} H^*$ with $\sigma^{}_2$ being the second Pauli matrix) respectively denote the lepton and Higgs doublets, and the superscript $c$ denotes the charge conjugated fields. And $Y^{}_\nu$, $M^{}_{\rm R}$ and $\mu^{}_{\rm s}$ are Yukawa and mass matrices in the flavor space. After the Higgs field acquires its vacuum expectation value $v=174$ GeV, the Yukawa couplings will lead to some mass terms connecting the left-handed neutrinos $\nu^{}_\alpha$ and RHNs: $(M^{}_{\rm D})^{}_{\alpha I} = (Y^{}_{\nu})^{}_{\alpha I} v$. In the $(\nu_{\alpha}^c, N^{}_{I}, S_{I}^c)$ basis, the complete neutrino mass matrix is thus given by
\begin{eqnarray}
M^{}_{\nu NS} = \begin{pmatrix}
0 & M^{}_{\rm D} & 0 \\
M^{T}_{\rm D} & 0 & M^{}_{\rm R} \\
0 & M^{T}_{\rm R} & \mu^{}_{\rm s} \\
\end{pmatrix} \;.
\label{1.3}
\end{eqnarray}
In this model the lepton number is violated by the Majorana mass terms of $S^{}_I$, so the $\mu^{}_{\rm s}$ parameters can be naturally small in the sense of $'$t Hooft \cite{Hooft} (i.e., the lepton number symmetry would get restored in the limit of $\mu^{}_{\rm s} \rightarrow 0$). Under the natural condition of $\mu^{}_{\rm s} \ll M^{}_{\rm D} \ll M^{}_{\rm R}$, the light neutrino masses can be derived by performing a block diagonalization of Eq.~(\ref{1.3}):
\begin{equation}
M^{}_{\nu} = M^{}_{\rm D} (M^{T}_{\rm R})^{-1}\mu^{}_{\rm s} M^{-1}_{\rm R} M^{T}_{\rm D} \;.
\label{1.4}
\end{equation}
We see that the smallness of neutrino masses follows naturally from small values of $\mu^{}_{\rm s}$, rendering superheavy RHNs no longer necessary. This can be intuitively understood from the following numerical relation
\begin{equation}
\left(\frac{M_\nu}{0.1~\mbox{eV}}\right) = \left(\frac{M_{\rm D}}{100~\mbox{GeV}}\right)^{2} \left(\frac{\mu^{}_{\rm s}}{1~\mbox{keV}}\right) \left(\frac{M_{\rm R}}{10~\mbox{TeV}}\right)^{-2} \;,
\label{1.5}
\end{equation}
which shows that the RHN mass scale can be naturally lowered to the TeV range for reasonable values of $\mu^{}_{\rm s}$ and $M^{}_{\rm D}$. Therefore, the ISS model provides an attractive framework that can naturally explain the smallness of neutrino masses while accommodating some sterile neutrinos potentially accessible at present or future experiments (see, e.g., Refs.~\cite{test1, test2}).

On the other hand, the RHN and $S^{}_I$ singlets (will be collectively referred to as sterile neutrinos) can be reorganized into pseudo-Dirac (PD) sterile neutrino pairs \cite{pseudo1}-\cite{pseudo4}. While a Dirac sterile neutrino can be thought of as two exactly degenerate Majorana sterile neutrinos with opposite CP phases (which will be realized for $\mu^{}_{\rm s} =0$), a PD sterile neutrino pair contains two nearly degenerate Majorana mass eigenstates with a tiny mass splitting (which will be realized for small values of $\mu^{}_{\rm s}$). To be concrete, in the one-generation sterile neutrino scenario (with only a pair of $N^{}_I$ and $S^{}_I$), one will arrive at a pair of PD sterile neutrinos with the masses $M^{}_{\rm R} + \mu^{}_{\rm s}/2$ and $M^{}_{\rm R} - \mu^{}_{\rm s}/2$. When it is generalized to the multi-generation sterile neutrino scenario (with multiple pairs of $N^{}_I$ and $S^{}_I$), one will similarly arrive at multiple pairs of PD sterile neutrinos which have their respective central masses and mass splittings.

At first sight, it seems that the quasi-degeneracy between a pair of PD sterile neutrinos is advantageous to the realization of resonant leptogenesis \cite{resonant1, resonant2}, which is expected to help us achieve a successful leptogenesis even with TeV scale sterile neutrinos. However, for generic ISS models with hierarchical PD sterile neutrino pairs (in which case the resonance effects for leptogenesis only take place within each pair of PD sterile neutrinos, in contrast with the scenario we will consider in this paper), the detailed study shows that the generated baryon asymmetry is smaller than its desired value by at least a few orders of magnitude \cite{us}. This is mainly due to (i) partial cancellation of lepton asymmetries within each pair of PD sterile neutrinos and (ii)
huge washout effects. In the literature, there already exist some studies on rescuing leptogenesis in the ISS model. For example, Refs.~\cite{rescue1, rescue2} have made use of the ARS mechanism (i.e., oscillations of sterile neutrinos) \cite{ARS1, ARS2} to do so. And Ref.~\cite{rescue3} has considered a non-standard cosmological expansion history and a quasi-degenerate mass spectrum of sterile neutrinos to do so.

Ref.~\cite{Agashe:2018cuf} has pointed out that quasi-degeneracies among different PD pairs could potentially enhance leptogenesis in the ISS framework, motivating us to examine whether such structures can arise naturally from an underlying flavor principle. In this paper, we investigate rescuing leptogenesis in the ISS model with the help of non-Abelian flavor symmetries which are motivated by the observed peculiar neutrino mixing pattern (see the next paragraph).
Unlike previous studies such as Ref.~\cite{rescue3}, where the nearly degenerate PD sterile neutrinos are effectively introduced through an arbitrary choice of their mass parameters, in our setup the required mass degeneracies emerge naturally as a consequence of the non-Abelian flavor symmetry. These symmetries enforce a highly constrained texture for $M^{}_{\rm R}$ (see Eq.~(\ref{1.7})), thereby providing a theoretical origin for the exact degeneracy among the three PD pairs and simultaneously reducing the number of free parameters.

Historically, the facts that the neutrino mixing angles $\theta^{}_{12}$ and $\theta^{}_{23}$ are close to some special values (i.e., $\sin^2 \theta^{}_{12} \sim 1/3$ and $\sin^2 \theta^{}_{23} \sim 1/2$, see the global-fit results for the neutrino-oscillation experiments in Refs.~\cite{global1, global2}) and the smallness of $\theta^{}_{13}$ suggest that the neutrino mixing matrix approximates to a very special form (which is referred to as the tribimaximal (TBM) mixing \cite{TB1,TB2}) as
\begin{eqnarray}
U^{}_{\rm TBM}= \displaystyle \frac{1}{\sqrt 6} \left( \begin{array}{ccc}
-2 & \sqrt{2} & 0 \cr
1 &  \sqrt{2}  & -\sqrt{3}  \cr
1 &  \sqrt{2}  & \sqrt{3} \cr
\end{array} \right)  \;,
\label{1.6}
\end{eqnarray}
which naturally predicts $\sin^2 \theta^{}_{12} = 1/3$, $\sin^2 \theta^{}_{23} = 1/2$ and $\theta^{}_{13} =0$. Such a remarkably simple and compact neutrino mixing pattern has attracted people to believe that there exists some flavor symmetry in the neutrino sector and have made a lot of attempts to explore the possible flavor symmetries (in particular non-Abelian flavor symmetries such as the ${\rm A}^{}_{4}$ and ${\rm S}^{}_{4}$ groups) underlying the observed neutrino mixing pattern \cite{FS1}-\cite{FS5}.
After the experimental determination of a relatively large $\theta^{}_{13}$, the TM1 (TM2) mixing \cite{TM-1}-\cite{TM-5} which retains the first (second) column of the TBM mixing while accommodating a non-zero $\theta^{}_{13}$ has become popular.

To be concrete, we propose a scenario in which mass degeneracies exist among different PD sterile neutrino pairs, which are then broken in a proper way. This configuration allows for resonant leptogenesis among different PD sterile neutrino pairs, potentially enhancing the generated baryon asymmetry. To achieve this, we utilize the above-mentioned non-Abelian flavor symmetries, which can naturally enforce mass degeneracies by unifying three RHNs (and also three $S^{}_I$ singlets) into a triplet representation (see, e.g., Refs.~\cite{S4-1}-\cite{MFS}). These symmetries ensure that the mass matrix $M^{}_{\rm R}$ in Eq.~(\ref{1.3}) adopts a symmetric form as given by
\begin{eqnarray}
M^{}_{\rm R} =M^{}_0 \begin{pmatrix}
1 & 0 & 0 \\
0 & 0 & 1 \\
0 & 1 & 0 \\
\end{pmatrix} \;.
\label{1.7}
\end{eqnarray}
In order to achieve the desired particular neutrino mixing pattern such as TM1 (TM2), at least one of the mass matrices $M^{}_{\rm D}$ and $\mu^{}_{\rm s}$ must encode a non-trivial flavor structure. A minimal choice (see also Ref.~\cite{MFS}) is that the mass matrix $\mu^{}_{\rm s}$ shares a same structure as $M^{}_{\rm R}$ in Eq.~(\ref{1.7}):
\begin{eqnarray}
\mu^{}_{\rm s}=\mu^{}_0 \begin{pmatrix}
1 & 0 & 0 \\
0 & 0 & 1 \\
0 & 1 & 0 \\
\end{pmatrix} \;,
\label{1.8}
\end{eqnarray}
while $M^{}_{\rm D}$ encodes a non-trivial flavor structure. For the sake of clarity, the structure of $M^{}_{\rm D}$ is taken to have the following form:
\begin{eqnarray}
M^{}_{\rm D} = \left( \begin{array}{ccc} a & 2 \,b & 2 \,c \\
2 \,b &  4 \,b + d & a - b - c - d \\
2 \,c & a - b - c - d & 4 \,c + d \\
      \end{array}\right)  \;,
\label{1.9}
\end{eqnarray}
which will naturally realize the popular TM1 neutrino mixing. With the help of Eq.~(\ref{1.4}), utilizing the light neutrino mass matrix which can be partially reconstructed in terms of the observed neutrino mixing angles and neutrino mass squared differences, we can derive the $M^{}_{\rm D}$ matrix accordingly. This derivation allows us to proceed with the subsequent calculations of baryon asymmetry.

It is easy to see that, for $M^{}_{\rm R}$ in Eq.~(\ref{1.7}) and $\mu^{}_{\rm s}$ in Eq.~(\ref{1.8}), the masses of three PD sterile neutrino pairs are completely degenerate, with each pair splitted by $\mu^{}_0$:
\begin{eqnarray}
M^{}_1 = M^{}_2 = M^{}_3 \simeq M^{}_0 - \frac{1}{2} \mu^{}_0 \;, \nonumber \\
M^{}_4 = M^{}_5 = M^{}_6 \simeq M^{}_0 + \frac{1}{2} \mu^{}_0 \;.
\label{1.10}
\end{eqnarray}
Then, as explained above, these degeneracies need to be broken in order for resonant leptogenesis among different PD sterile neutrino pairs to come into play. In this work, we consider the following two well-motivated approaches to generate such desired mass splittings: (i) they can be naturally induced by the renormalization group evolution (RGE) effects; (ii) when the mass matrix $\mu^{}_{\rm s}$ also contains some non-trivial flavor structure parallel to that in Eq.~(\ref{1.9}) (so that the TM1 mixing scheme will be preserved), the flavor non-universality of $\mu^{}_{\rm s}$ will break the mass degeneracies among the three PD sterile neutrino pairs. By analyzing these two scenarios, we aim to assess the viability of leptogenesis in the ISS model equipped with non-Abelian flavor symmetries, and to identify the conditions under which the observed baryon asymmetry of the Universe can be successfully reproduced.

The remaining parts of this paper are organized as follows. In the next section, we recapitulate some basic formulas of ISS model and leptogenesis as a basis of our study. In sections~3 and 4,
we separately investigate the potential for leptogenesis within the ISS model equipped with non-Abelian flavor symmetries under the two distinct scenarios outlined above. For these scenarios, we will also investigate the consequences of the model with respect to the charged lepton flavor violation processes. Finally, the summary of our main results will be given in section~5.

\section{Some basic formulas of ISS model and leptogenesis}

In this section, we recapitulate some basic formulas of ISS model and leptogenesis as a basis of our study.

In order to facilitate the leptogenesis calculations, it is most convenient to work in the sterile neutrino mass basis, where the bottom-right sub-matrix of Eq.~(\ref{1.3}) (denoted as $ M^{}_{NS}$) become real, positive and diagonal. For this purpose, one can diagonalize $ M^{}_{NS}$
via the unitary transformation $V^{}_{\rm R}$ \cite{Diag}
\begin{eqnarray}
V^{T}_{\rm R} M^{}_{NS} V^{}_{\rm R} \approx
\begin{pmatrix}
V^{T}_{1}\left[-M^{}_{\rm R}+\frac{1}{2}  \mu^{}_{\rm s} \right] V^{}_{1} &  0  \\
0	&  V^{T}_{2}\left[M^{}_{\rm R}+\frac{1}{2}   \mu^{}_{\rm s} \right] V^{}_{2}
\end{pmatrix} \;,
\label{2.1}
\end{eqnarray}
with
\begin{eqnarray}
V^{}_{\rm R}\approx\frac{1}{\sqrt{2}}
\begin{pmatrix}
\mathbf{1}+\frac{\mu^{}_{\rm s} {M^{-1}_{\rm R}}^{}}{4}	& \mathbf{1}-\frac{\mu^{}_{\rm s} {M^{-1}_{\rm R}}^{}}{4} \\
-\mathbf{1}+\frac{\mu^{}_{\rm s} {M^{-1}_{\rm R}}^{}}{4} & \mathbf{1}+\frac{\mu^{}_{\rm s} {M^{-1}_{\rm R}}^{}}{4}
\end{pmatrix}
\begin{pmatrix}
V^{}_{1} & 0 \\
0  & V^{}_{2}
\end{pmatrix}  \;,
\label{2.2}
\end{eqnarray}
where $V^{}_{1}$ and $V^{}_{2}$ are the unitary matrices for diagonalizing $-M^{}_{\rm R}+1/2  \mu^{}_{\rm s}$ and $M^{}_{\rm R}+1/2  \mu^{}_{\rm s}$, respectively.
In this way the complete neutrino mass matrix is transformed to the following form
\begin{eqnarray}
M_{\nu NS}^{\prime} \simeq \begin{pmatrix}
0 & v Y^{\prime}_{\rm D} & v Y^{\prime}_{\rm L} \\
v \left(Y^{\prime}_{\rm D}\right)^{T} & V^{T}_{1}\left[-M^{}_{\rm R}+\frac{1}{2}  \mu^{}_{\rm s} \right] V^{}_{1} & 0 \\
v \left(Y^{\prime}_{\rm L}\right)^{T} & 0 & V^{T}_{2}\left[M^{}_{\rm R}+\frac{1}{2}   \mu^{}_{\rm s} \right] V^{}_{2} \end{pmatrix} \;,
\label{2.3}
\end{eqnarray}
where $Y^{\prime}_{\rm D}$ and $Y^{\prime}_{\rm L}$ correspond to the rotated $3\times 3$ Yukawa matrices that couple the sterile neutrinos to the SM leptons, and they can be analytically expressed as follows \cite{Yv}
\begin{eqnarray}
Y^{\prime}_{\rm D} & \simeq & \frac{1}{\sqrt{2}\,v}\left[ M^{}_{\rm D} \left( \mathbf{1} + \frac{\mu^{}_{\rm s} {M^{-1}_{\rm R}}}{4} \right) V^{}_{1} \right] \;,
\nonumber\\
Y^{\prime}_{\rm L} & \simeq & \frac{1}{\sqrt{2}\,v}\left[ M^{}_{\rm D} \left( \mathbf{1} - \frac{\mu^{}_{\rm s} {M^{-1}_{\rm R}}}{4} \right)V^{}_{2} \right] \;.
\label{2.4}
\end{eqnarray}

In the low-scale (TeV or so) leptogenesis regime which we are interested in, all the three lepton flavors have become distinguishable from one another and the lepton asymmetries stored in them should be tracked separately \cite{flavor1, flavor2}. On the other hand, the contributions of the nearly degenerate sterile neutrinos to the final baryon asymmetry are on the same footing and should be taken into consideration altogether. For these two reasons, the final baryon asymmetry can be obtained as
\begin{eqnarray}
Y^{}_{\rm B}  = c Y^{}_{\rm L} = c r \sum^{}_\alpha \varepsilon^{}_{\alpha} \kappa^{}_\alpha \;,
\label{2.5}
\end{eqnarray}
where $c \simeq -1/3$ is the conversion efficiency from the lepton asymmetry to the baryon asymmetry via the sphaleron processes, $r \simeq 4 \times 10^{-3}$ measures the ratio of the equilibrium number density of sterile neutrinos to the entropy density, and $\varepsilon^{}_\alpha$ is a sum over the sterile neutrinos (i.e., $\varepsilon^{}_\alpha= \sum^{}_{I}\varepsilon^{}_{I \alpha}$) of the CP asymmetries for their decays:
\begin{eqnarray}
	\varepsilon^{}_{I \alpha } \equiv \frac{ \Gamma \left( N^{}_I \to L^{}_\alpha + H \right) - \Gamma \left( N^{}_I \to \overline{L}^{}_\alpha + \overline{H} \right) }
	{\sum^{}_{\alpha} \Gamma \left( N^{}_I \to L^{}_\alpha + H \right) + \Gamma \left( N^{}_I \to \overline{L}^{}_\alpha + \overline{H} \right)  } \;.
\label{2.6}
\end{eqnarray}
Finally, $\kappa^{}_\alpha$ are the efficiency factors that take account of the washout effects due to the inverse decays of sterile neutrinos and various lepton-number-violating scattering processes. The explicit expressions for $\varepsilon^{}_{I \alpha}$ and $\kappa^{}_\alpha$ are given below.

In the resonant leptogenesis regime relevant for our study, which involves three pairs of PD sterile neutrinos (i.e., six nearly degenerate states), the CP asymmetries differ from the conventional two-state expressions. In particular, the effective neutrino Yukawa couplings in our framework are given by the following form \cite{Pilaftsis:2003gt}-\cite{daSilva:2022mrx}
\begin{eqnarray}
\left( \bar{h}^{}_{+} \right)^{}_{\alpha I } = h_{\alpha I } + {\rm i} \mathcal{V}^{}_{\alpha I }
	- {\rm i} \sum_{J, K, L, M, N=1}^{6} h^{}_{\alpha J }  \mathcal{F}^{}_{IJKLMN}  \;,
\label{2.7}
\end{eqnarray}
where $h^{}_{}= (Y^{\prime}_{\rm D}, Y^{\prime}_{\rm L})$, and
\begin{eqnarray}
 	&&\hspace{-1.5cm} \mathcal{V}^{}_{\alpha I } =-\sum^{}_{\beta =e, \mu, \tau} \sum^{}_{J \neq I} \frac{h^{*}_{ \beta I } h^{}_{\beta J } h^{}_{\alpha J } }{16 \pi} f\left( \frac{M^2_{J}}{M^2_{I}} \right)  \;, \nonumber \\
	&&\hspace{-1.5cm} \mathcal{F}^{}_{IJKLMN} =\frac{1}{\left( M^{2}_{I} - M^{2}_{J} + 2 {\rm i} M^{2}_{I} A^{}_{JJ} \right) F^{(4)}_{IJKLMN}}  \times
\{ M^{}_{I}  \left( \mathcal{M}^{}_{IIJ } + \mathcal{M}^{}_{JJI} \right) \nonumber \\
	&&\hspace{0.6cm} - {\rm i} R^{}_{IN} \left[ \mathcal{M}^{}_{INJ } \left( \mathcal{M}^{}_{IIN } + \mathcal{M}^{}_{NNI } \right) + \mathcal{M}^{}_{JJN} \left( \mathcal{M}^{}_{INI } + \mathcal{M}^{}_{NIN} \right) \right]  \nonumber \\
	&&\hspace{0.6cm} -{\rm i} R^{}_{IM} \frac{1}{F^{(1)}_{IMN}} \left[ \mathcal{M}^{}_{IMJ } \left( \mathcal{M}^{}_{IIM } + \mathcal{M}^{}_{MMI } \right) + \mathcal{M}^{}_{JJM} \left( \mathcal{M}^{}_{IMI } + \mathcal{M}^{}_{MIM} \right) \right]  \nonumber \\
	&&\hspace{0.6cm} -{\rm i} R^{}_{IL} \frac{1}{F^{(2)}_{ILMN}} \left[ \mathcal{M}^{}_{ILJ } \left( \mathcal{M}^{}_{IIL } + \mathcal{M}^{}_{LLI } \right) + \mathcal{M}^{}_{JJL} \left( \mathcal{M}^{}_{ILI } + \mathcal{M}^{}_{LIL} \right) \right]  \nonumber \\
	&&\hspace{0.6cm} -{\rm i} R^{}_{IK} \frac{1}{F^{(3)}_{IKLMN}} \left[ \mathcal{M}^{}_{IKJ } \left( \mathcal{M}^{}_{IIK } + \mathcal{M}^{}_{KKI } \right) + \mathcal{M}^{}_{JJK} \left( \mathcal{M}^{}_{IKI } + \mathcal{M}^{}_{KIK} \right) \right]
\} \;,
	\label{2.8}
\end{eqnarray}
with $f(x) = \sqrt{x} [ 1 - (1+x) \ln (1+1/x)]$ being the loop function, $\mathcal{M}^{}_{IJK} \equiv M^{}_{I} A^{}_{JK}$ and
\begin{eqnarray}
&& A^{}_{IJ} = \frac{ \left( h^{\dagger} h \right)^*_{IJ} }{16 \pi} \;, \hspace{1cm} R^{}_{IJ} = \frac{ M^{2}_{I} }{ M^{2}_{I} - M^{2}_{J} + 2 {\rm i} M^{2}_{I} A^{}_{JJ} }  \;,\nonumber \\
&& F^{(1)}_{IJK}= 1+ 2 {\rm i} {\rm Im} (R^{}_{IK}) T^{}_{IJK}\;,\nonumber \\
&& F^{(2)}_{IJKL}= F^{(1)}_{IJL}+ 2 {\rm i} {\rm Im} \left(R^{}_{IK} \frac{1}{F^{(1)}_{IKL}}\right) T^{}_{IJK}\;,\nonumber \\
&& F^{(3)}_{IJKLM}= F^{(2)}_{IJLM}+ 2 {\rm i} {\rm Im} \left(R^{}_{IK} \frac{1}{F^{(2)}_{IKLM}}\right) T^{}_{IJK}\;,\nonumber \\
&& F^{(4)}_{IJKLMN}= F^{(3)}_{IJLMN}+ 2 {\rm i} {\rm Im} \left(R^{}_{IK} \frac{1}{F^{(3)}_{IKLMN}}\right) T^{}_{IJK}\;,\nonumber \\
&& T^{}_{IJK}= \frac{ M^{2}_{I} |A^{}_{JK}|^{2} + M^{}_{J}M^{}_{K}{\rm Re} A^{2}_{JK}}{ M^{2}_{I} - M^{2}_{J} + 2 {\rm i} M^{2}_{I} A^{}_{JJ} } \;.
\label{2.9}
\end{eqnarray}
The corresponding CP-conjugate effective Yukawa couplings $\left( \bar{h}^{}_{-} \right)^{}_{\alpha I }$ can be obtained by simply making the replacements $h^{}_{\alpha I } \to h^{*}_{\alpha I }$ in the above expressions. Furthermore, in terms of these effective Yukawa couplings, the decay widths of sterile neutrinos can be expressed as
\begin{eqnarray}
\Gamma \left( N^{}_I \to L^{}_{\alpha} + H \right) = \frac{ M^{}_{I} }{16\pi} \left| \left( \bar{h}^{}_{+} \right)^{}_{\alpha I } \right|^2 \;, \hspace{1cm}
\Gamma \left( N^{}_I \to \overline{L}^{}_\alpha + \overline{H} \right) = \frac{ M^{}_{I} }{16\pi} \left| \left( \bar{h}^{}_{-} \right)^{}_{\alpha I } \right|^2 \;,
\label{2.10}
\end{eqnarray}
Therefore, the flavor-specific CP asymmetries for the decays of the $I$-th sterile neutrino are given by
\begin{eqnarray}
	\varepsilon^{}_{I \alpha }
	= \frac{ \left| (\bar{h}^{}_{+})^{}_{\alpha I } \right|^2 - \left| (\bar{h}^{}_{-})^{}_{\alpha I } \right|^2  }
	{ \left( \bar{h}^{\dagger}_{+} \bar{h}^{}_{+} \right)^{}_{I I} + \left( \bar{h}^{\dagger}_{-} \bar{h}^{}_{-} \right)^{}_{II} } \;.
\label{2.11}
\end{eqnarray}

In the standard seesaw models, the interactions of the unstable heavy neutrinos can be effectively described within the framework of $\Delta L=2$ scattering processes mediated by on-shell heavy neutrinos $L_\alpha  \Phi \leftrightarrow  L_\beta^C  \Phi^\dagger$. These scatterings can be separated into on-shell and off-shell contributions, where the former yields the inverse decay process. The corresponding reaction density is given by
\begin{eqnarray}
\gamma_{\Delta L=2, I\alpha }^{\rm on} = \frac{\gamma^D_{ I\alpha }}{4}  \equiv
\frac{1}{4}\, n_{N_I}^{\rm
eq}\,\frac{\mathcal{K}_{1}}{\mathcal{K}_2}
\,\Gamma_{ I \alpha} =
\frac{4}{z} \frac{M_I^4}{246\pi^3}(|(\bar{h}^{}_{+})_{\alpha I }|^2+|(\bar{h}^{}_{-})_{\alpha I }|^2)
\mathcal{K}_{1}(z\, M_I/M_1) \;,
\label{2.12}
\end{eqnarray}
where $\mathcal{K}_{1,2}(z)$ are the modified Bessel functions, $z\equiv M_{1}/T$ with $T$ being the thermal bath temperature, and $\Gamma^{}_{ I\alpha }=\Gamma \left( N^{}_I \to L^{}_{\alpha} + H \right)+\Gamma \left( N^{}_I \to \overline{L}^{}_\alpha + \overline{H} \right)$. The strength of washout effects due to inverse decays is controlled by the washout parameters
\begin{eqnarray}
K^{}_{\alpha}= \sum^{}_{I} K_{ I \alpha} =  \sum^{}_{I} \frac{\Gamma^{}_{ I \alpha}}{H(T=M^{}_{I})} \;,
\label{2.13}
\end{eqnarray}
where $H(T) =1.66 \sqrt{g_*}\, T^2/M^{}_{\rm Pl}$ is the Hubble rate, $g^{}_*$ the number of relativistic degrees of freedom and $M^{}_{\rm Pl} = 1.22\times 10^{19}$ GeV the Planck mass. But in the ISS model, it is natural to expect that the lepton-number-violating washout will go to zero in the limit of vanishing $\mu^{}_{\rm s}$. And it has been shown that the suppression of the washout proceeds through the destructive interference of one member of a PD sterile neutrino pair with the other, leaving a washout which vanishes in the lepton-number-conserving (LNC) limit (i.e., $\mu^{}_{\rm s}=0$) \cite{Yv, ISSSO10}.
The total washout is protected by the symmetry and vanishes exactly in the LNC limit, regardless of the large values in the naive parameters of Eq.~(\ref{2.13}). The introduction of a sizable $\mu_{\rm s}$ breaks this protective symmetry, breaking the cancellation and restoring the standard inverse-decay washout. This highlights the critical role of interference effects, which are absent in standard seesaw models but paramount in approximate LNC frameworks like the ISS model. Following the treatment proposed in Ref.~\cite{Yv}, one must return to the fundamental definition of $\Delta L = 2$ scattering reaction density
\begin{eqnarray}
 \gamma_{\Delta L=2,\alpha}^{\rm tot}=
\frac{1}{z}\frac{M_{1}^{4}}{64\,\pi^{4}}\,\int_{x_{\rm
thr}}^{\infty}dx\,\sqrt{x}\,\hat{\sigma}_{\Delta
L=2,\alpha}(x)\,\mathcal{K}_{1}(z\,\sqrt{x})\;,
\label{2.14}
\end{eqnarray}
where $x\equiv s/M_{N_1}^2$, and $x^{}_{\rm  thr}$ denotes the kinematic threshold. The $s$-channel reduced cross section is given by \cite{Pilaftsis:2005rv}
\begin{eqnarray}
\widehat{\sigma}^{}_{\Delta L=2,\alpha}(x)
\!\!&=&\!\!  \sum_{\beta=1}^{3}\ \sum_{I, J=1}^{6}\ {\rm Re}\, \Bigg\{\,
\bigg[\,(\bar{h}^{}_+)^*_{\alpha I}\,(\bar{h}_+)_{\alpha J}\,
(\bar{h}^{}_+)^*_{\beta I}\,(\bar{h}_+)_{\beta J}\: +\:
(\bar{h}^{}_-)^*_{\alpha I}\,(\bar{h}_-)_{\alpha J}\,
(\bar{h}^{}_-)^*_{\beta I}\,(\bar{h}_-)_{\beta J}\,\bigg]
\:\frac{x\,\sqrt{a_{I}a_{J} } }{4\pi P^{*}_{I}P_{J} }\ \Bigg\} \;,\nonumber \\
\label{2.15}
\end{eqnarray}
where $P_{I}^{-1}= \left(x-a_{I}+i\sqrt{a_{I} c_{I}}\right)^{-1}$ is the Breit-Wigner $s$-channel
propagator, with $a_{I}=(M_{I}/M_{1})^{2}$ and $c_{I}=(\Gamma_{I}/M_{1})^{2}$.
A key advantage of this expression is that the cross section naturally incorporates the interference terms among different heavy neutrinos, thereby enabling a consistent definition of the effective washout parameter as
\begin{eqnarray}
K^{\rm eff}_{\alpha} \equiv K^{}_{\alpha} \cdot  \frac{\gamma_{\Delta L=2,\alpha}^{\rm tot}}{\gamma^D_{ \alpha}/4} \;,
\label{2.16}
\end{eqnarray}
where $\gamma^{D}_{\alpha}\equiv \sum_{I}^{6} \gamma^{D}_{ I \alpha}/6$.

In most of realistic leptogenesis parameter space, one has $K^{\rm eff}_{\alpha} \geqslant 1$, known as the strong washout regime. In this regime, the efficiency factors are roughly inversely proportional to the washout parameter, as described by
\begin{eqnarray}
\kappa^{}_{\alpha} \approx \frac{2}{z^{}_{B} K^{\rm eff}_{\alpha}}\left[1-e^{-\frac{1}{2}z^{}_{B} K^{\rm eff}_{\alpha}}\right] \;,
\label{2.16}
\end{eqnarray}
where $z_{B}$ is well approximated by \cite{zB}
\begin{eqnarray}
z_{B} \approx  1+\frac{1}{2}\ln\left[1+\frac{\pi {K^{\rm eff}_{\alpha}}^2}{1024}\left(\ln\frac{3125\pi {K^{\rm eff}_{\alpha}}^2}{1024}\right)^{5}\right] \;.
\label{2.17}
\end{eqnarray}
For completeness, we also consider the weak washout regime where $K^{\rm eff}_{\alpha} < 1 $. In this regime, depending on whether $K^{}_{\alpha}$ is larger or smaller than 1, there are the following two distinct cases
\begin{eqnarray}
\kappa_{\alpha} \approx
\left\{
\begin{array}{ll}
\frac{1}{4} K^{\rm eff}_{\alpha} \left(\frac{3\pi}{2} - z^{}_{B}\right) \hspace{0.85cm} (K^{}_{\alpha} > 1) \;; \\
\frac{9\pi^{2}}{64} K^{\rm eff}_{\alpha} K^{}_{\alpha} \hspace{1.8cm} (K^{}_{\alpha} < 1) \;.
\end{array}
\right.
\label{2.18}
\end{eqnarray}

Finally, with the help of Eq.~(\ref{2.11}), we show that for the ISS model with $M^{}_{\rm R}$ in Eq.~(\ref{1.7}) and $\mu^{}_{\rm s}$ in Eq.~(\ref{1.8}) the total CP asymmetries (summed over the sterile neutrinos) are greatly suppressed. Let us take the first PD pair of sterile neutrinos (which contains the first and fourth sterile neutrinos) as an example: the relation $V^{}_{1}={\rm i} V^{}_{2}$ holds in the present scenario (see Eq.~(\ref{2.1})), and it implies
\begin{eqnarray}
&& h^{}_{\alpha 1 }=i h^{}_{\alpha 4 }\;,
\hspace{0.5cm} \mathcal{V}^{}_{\alpha 1 }=i \mathcal{V}^{}_{\alpha 4 }\;, \hspace{0.5cm}
(\mathcal{B}_{+})_{\alpha 1 }\simeq i (\mathcal{B}_{-})^{*}_{ \alpha 4}\;,\hspace{0.5cm}
(\mathcal{B}_{-})_{\alpha 1 }\simeq i (\mathcal{B}_{+})^{*}_{\alpha 4 }\;,
\label{2.19}
\end{eqnarray}
where $(\mathcal{B}_{+})_{\alpha I }=\sum_{J, K, L, M, N=1}^{6} h^{}_{\alpha J }  \mathcal{F}^{}_{IJKLMN}$  (while $(\mathcal{B}_{-})_{\alpha I }$ can be obtained by making the replacement $h^{}_{\alpha I } \to h^{*}_{\alpha I }$). It follows that
\begin{eqnarray}
 \left| (\bar{h}^{}_{+})^{}_{\alpha 1 } \right|^2 \approx \left| (\bar{h}^{}_{-})^{}_{\alpha 4 } \right|^2 \;,
 \hspace{0.5cm}
 \left| (\bar{h}^{}_{-})^{}_{ \alpha 1} \right|^2 \approx \left| (\bar{h}^{}_{+})^{}_{\alpha 4 } \right|^2 \;,
\label{2.20}
\end{eqnarray}
and consequently $\varepsilon^{}_{ 1\alpha}$ and $\varepsilon^{}_{4\alpha}$ cancel each other out to the leading order. Therefore, the total CP asymmetry for the first PD pair of sterile neutrinos is greatly suppressed. And the same is true for the other two pairs of PD sterile neutrinos. This is why we need to break the degeneracies among different PD sterile neutrino pairs in order to potentially achieve sufficient baryon asymmetry.

\section{Renormalization group evolution assisted leptogenesis}

In this section, we study the scenario that the desired mass splittings among different PD sterile neutrino pairs are generated from the RGE effects. In this scenario, as we will see, it is the differences among the Yukawa couplings of different RHNs that induce mass splittings among different PD sterile neutrino pairs.

\subsection{Consequence for leptogenesis}

In the literature, the flavor symmetries are usually placed at very high energy scales $\Lambda^{}_{\rm FS}$ \cite{FS2}-\cite{FS3}. When dealing with leptogenesis which takes place around the sterile neutrino mass scale $M^{}_0$, the RGE effects should be taken into account if there is a large gap between $\Lambda^{}_{\rm FS}$ and $M^{}_0$. In the SM framework, the one-loop RGE equations of the neutrino Yukawa coupling matrix $Y_\nu$ and the mass matrix $M^{}_{\rm R}$ are described by \cite{ISSRGE}
\begin{eqnarray}
&& 16\pi^2  \frac{{\rm d}Y^{}_\nu}{{\rm d}t}  =\left( \alpha^{}_\nu -\frac{3}{2} Y^{}_l Y^\dagger_l + \frac{3}{2} Y^{}_\nu Y^\dagger_\nu \right)Y^{}_\nu \; ,
 \nonumber \\
&& 16\pi^2  \frac{{\rm d}M^{}_{\rm R}}{{\rm d}t}  =  M^{}_{\rm R} Y^\dagger_\nu Y^{}_\nu  \; ,
\label{3.1}
\end{eqnarray}
where $t = \ln \left(\mu/ \Lambda^{}_{\rm FS}\right)$ has been defined with $\mu$ being the renormalization scale. The parameter $\alpha_\nu $ is defined as
\begin{eqnarray}
\alpha^{}_\nu  =  {\rm tr} \left(3 Y^{}_u Y^\dagger_u + 3 Y^{}_d Y^\dagger_d + Y^{}_l Y^\dagger_l +  Y^{}_\nu Y^\dagger_\nu \right)-\frac{9}{20} g^2_1 - \frac{9}{4} g^2_2  \;,
\label{3.2}
\end{eqnarray}
where $g^{}_1$ and $g^{}_2$ are the $\rm U\left(1\right)^{}_{\rm Y}$ and $\rm SU\left(2\right)^{}_{\rm L}$ gauge coupling constants, and $Y^{}_{u,d,l}$ are respectively the Yukawa matrices for up-type quarks, down-type quarks and charged leptons.

Before performing the numerical calculations, for the purpose of illustration, we take the diagonal terms of $M^{}_{\rm R}$ as an example to show the impacts of RGE effects on leptogenesis. For the $II$ element of $M^{}_{\rm R}$, an integration of Eq.~(\ref{3.1}) yields the following relation between its value at the flavor-symmetry scale $\Lambda^{}_{\rm FS}$ and mass scale $M^{}_0$:
\begin{eqnarray}
	M^{}_{II} \left( M^{}_0 \right) \simeq M^{}_{II} \left( \Lambda^{}_{\rm FS} \right)
	\left[1 - \frac{1}{16 \pi^2} \left(Y^{ \dagger}_{\nu} Y^{}_\nu \right)^{}_{II} \ln \left(  \frac{  \Lambda^{}_{\rm FS} }{ M^{}_0 }  \right) \right] \;,
\label{3.3}
\end{eqnarray}
which yields the following splitting between $M^{}_{II} \left( M^{}_0 \right)$ and $M^{}_{JJ} \left( M^{}_0 \right)$:
\begin{eqnarray}
	M^{}_{II} \left( M^{}_0 \right) - M^{}_{JJ} \left( M^{}_0 \right)
	\simeq \frac{ M^{}_{I} \left( \Lambda^{}_{\rm FS} \right) }{16 \pi^2}
	\left[ \left( Y^{ \dagger}_{\nu} Y^{}_\nu \right)^{}_{JJ} - \left( Y^{ \dagger}_{\nu} Y^{}_\nu \right)^{}_{II} \right]
	\ln \left( \frac{  \Lambda^{}_{\rm FS} }{ M^{}_0 } \right) \;.
\label{3.4}
\end{eqnarray}
This shows that it is really the differences among the Yukawa couplings of different RHNs that induce mass splittings among different PD sterile neutrino pairs, as mentioned above. One can also see that such mass splittings increase significantly as the mass scale $\mu^{}_0$ decreases (accompanying which the magnitudes of $Y^{}_\nu$ will increase as indicated by Eq.~(\ref{1.5})), potentially exceeding the scale of $\mu^{}_0$ itself.
At this point, although the Yukawa coupling matrix still satisfies the relationship described in Eq.~(\ref{2.19}), the $(\mathcal{B}_{\pm})_{\alpha I }$ terms no longer do so.
Consequently, the cancellation of CP asymmetries is weakened, allowing for the potential generation of a sufficiently large baryon asymmetry. Finally, we stress that the off-diagonal-term contributions are also significant in the scenario considered in this paper, and we have also taken account of them in the following calculations.

\begin{figure*}
\centering
\includegraphics[width=6.5in]{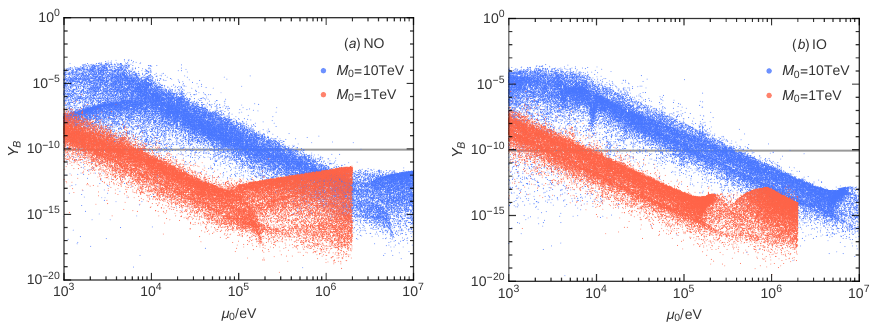}
\caption{ For the scenario studied in section~3.1, in the NO (a) and IO (b) cases, the allowed values of $Y^{}_{\rm B}$ as functions of $\mu_0$ for some benchmark values of $M^{}_0$. The horizontal line stands for the observed value of $Y^{}_{\rm B}$. }
\label{fig1}
\end{figure*}

Now, we are ready to perform the numerical calculations.
In Figure~\ref{fig1}(a) and (b) (for the normal ordering (NO) and inverted ordering (IO) cases of light neutrino masses, respectively) we have shown the allowed values of $Y^{}_{\rm B}$ as functions of $\mu^{}_0$ for the benchmark values of $M^{}_0 = 1$ and 10 TeV.
These results are obtained for the following parameter settings: for the neutrino mass squared differences and neutrino mixing angles, we employ the global-fit results for them \cite{global1, global2}. For the two Majorana CP phases of the neutrino mixing matrix, we allow them to vary in the range 0---$2\pi$, while the Dirac CP phase is constrained by the TM1 mixing scheme realized by the scenario considered in the present paper. The lightest neutrino mass ($m^{}_1$ or $m^{}_3$ in the NO or IO case) is varied in the range between $0.001$ eV and $0.1$ eV. For $\Lambda^{}_{\rm FS}$, we fix $\Lambda_{\text{FS}} / M^{}_0= 10^2$ as a benchmark value (in fact, the dependence of the final results on it is weak since the RGE effects only depend on it in a logarithmic manner).
To conform to the essential idea of the ISS model (i.e., $\mu^{}_{\rm s} \ll M^{}_{\rm D} \ll M^{}_{\rm R}$), in the calculations we have required $\mu^{}_{\rm s}$ to be at least two orders of magnitude smaller than $M^{}_{\rm D}$, and $M^{}_{\rm D}$ to be at least two orders of magnitude smaller than $M^{}_{\rm R}$.

The results show that in both the NO and IO cases the observed value of $Y^{}_{\rm B}$ has chance to be successfully reproduced. It is clear that for smaller values of $M^{}_0$ (e.g., $M^{}_0 = 1 $ TeV), relatively small values of $\mu^{}_0$ (around $10^3$ eV) are needed to reproduce the observed value of $Y^{}_{\rm B}$.
As $M^{}_0$ increases, the scale of $\mu^{}_0$ needed to reproduce the observed $Y^{}_{\rm B}$ also rises. In the figure, two turning points are observed. Upon further analysis, we find that for $M^{}_0 = 10 $ TeV, the left turning point results from a decrease in $\mu^{}_0$, leading to an increase in the sterile neutrino mass splittings induced by the RGE effects.
Once these splittings become too large, significant hierarchies among the three PD sterile neutrino pairs arise. In particular, around $\mu^{}_0 \sim 10^3$ eV, only one PD pair remains quasi-degeneracy (under the degeneracy criterion of $\Delta M/M \lesssim 10^{-3}$ adopted in this work), effectively reducing the system to a two-state resonance. As a result, the six-state framework underlying Eq.~(\ref{2.8}) is no longer realized. We emphasize that this conclusion is determined by the specific numerical threshold used to define quasi-degeneracy. Our stringent choice ensures that Eq.~(\ref{2.8}) only applies within a fully controlled six-state scenario, thereby placing the region near the left turning point outside the valid domain of our analysis.
Conversely, the right turning point occurs due to an increase in $\mu^{}_0$, which causes the sterile neutrino mass splittings induced by the RGE effects to approach the scale of $\mu^{}_0$. In this regime, $\mu^{}_0$ also contributes significantly and is equally important. In the six-state scenario with three PD pairs, the pattern of the resonant enhancement is more intricate than in the conventional two-state case. The right turning point appears when the mass splittings of each PD pair (controlled by $\mu_0$) become comparable to their decay widths, so that each PD pair can individually enter their own resonant regime. By contrast, the earlier enhancement region is dominated by the mass splittings among different PD pairs. It is worth noting that even in the standard two-state case, resonant leptogenesis does not merely feature a single resonance regime but can exhibit multiple enhancement regimes depending on the relative sizes of the mass splittings and decay widths.
Overall, it can be observed that as $M^{}_0$ varies in the range of 1 to 10 TeV, the observed value of $Y^{}_{\rm B}$ can be successfully reproduced across a wide scale of $\mu^{}_0$, spanning from $10^3$ to $10^6$ eV.

\subsection{Charged lepton flavour violation}

In the context of our model, the Yukawa interactions responsible for generating neutrino masses also give rise to charged lepton flavor violation (cLFV). Among various possible cLFV channels, the radiative decay $l_\alpha \to l_\beta \gamma $ is the most widely studied one, as it provides the most stringent experimental limits and is expected to exhibit the largest branching ratios. According to the MEG collaboration \cite{MEG:2016leq}, one now has the experimental limit ${\rm BR}(\mu \to e \gamma) < 4.2 \times 10^{-13}$. Therefore, the evaluation of branching ratios for these decays within our framework not only acts as a key consistency check of the model but also establishes a direct connection between the neutrino mass generation mechanism and experimentally accessible low-energy signals.

Within our framework, ${\rm BR}(l_\alpha \to l_\beta \gamma)$ can be calculated according to \cite{Forero:2011pc}
\begin{eqnarray}
{\rm BR}(l_\alpha \to l_\beta \gamma)= \frac{\alpha^3_W s_W^2}{256 \pi^2}\frac{m_{l_\alpha}^5}{M_W^4}
\frac{1}{\Gamma_{l_{\alpha}}}|G_{\alpha \beta}^W|^2 \; ,
\label{3.5}
\end{eqnarray}
where
\begin{eqnarray}
&& G_{\alpha \beta}^W=\sum_{i=1}^9 K^*_{\alpha i} K_{\beta i} G_\gamma^W \left(\frac{M^2_{i}}{M_W^2}\right) \; ,
 \nonumber \\
&& G_\gamma^W(x)=\frac{1}{12(1-x)^4}(10-43x+78x^2-49x^3+18x^3\ln{x}+4x^4) \;.
\label{3.6}
\end{eqnarray}
Here $ \alpha _W \equiv g_W^2/(4\pi)$ denotes the weak fine-structure constant, $s_W^2= \sin^2 \theta _W$ with $\theta _W$ being the Weinberg angle, $M_W$ the $W$-boson mass, $m_{l_\alpha}$ and $\Gamma_{l_{\alpha}}$ the mass and total decay width of the decaying charged lepton $l_\alpha$.
Finally, one has the matrix $K=\left(K_{L},K_{H}\right)$ with
\begin{eqnarray}
&& K_L = \left(I-\frac{1}{2}(M^{}_{\rm D}, 0)^*(M_{NS}^{\prime -1})^{*} M_{NS}^{\prime -1} (M^{}_{\rm D}, 0)^T\right) U^{\prime *} \; ,
 \nonumber \\
&& K_H = (M^{}_{\rm D}, 0)^* (M_{NS}^{\prime -1})^{*} V \;,
\label{3.7}
\end{eqnarray}
where $M_{NS}^{\prime}$ is the sterile-neutrino mass matrix after taking account of the RGE effects.

\begin{figure*}
\centering
\includegraphics[width=6.5in]{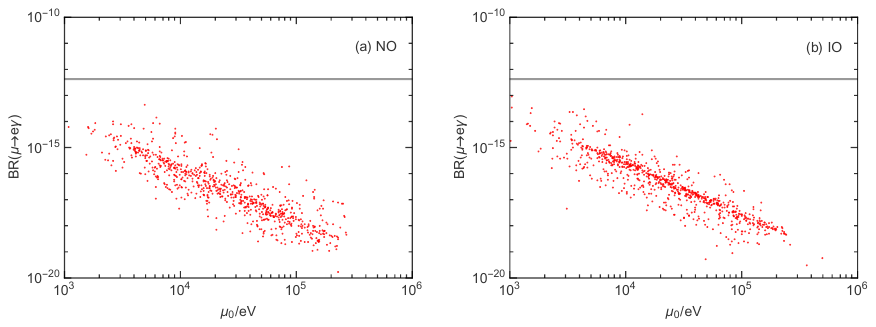}
\caption{ For the scenario studied in section~3.1, in the parameter space that allows for a reproduction of the observed value of $Y^{}_{\rm B}$, the allowed values of ${\rm BR}(\mu \to e \gamma)$ as functions of $\mu_0$ in the NO (a) and IO (b) cases. The horizontal line stands for the current upper bound on ${\rm BR}(\mu \to e \gamma)$. }
\label{fig2}
\end{figure*}

Now we investigate the consequences of the model with respect to the $\mu \to e \gamma $ process. In Figure~\ref{fig2}(a) and (b) (for the NO and IO cases, respectively), we have shown the allowed values of ${\rm BR}(\mu \to e \gamma)$ as functions of $\mu_0$. With the same parameter settings as in section~3.1, these results are obtained within the parameter space that allows for a successful reproduction of the observed baryon asymmetry of the Universe, with $M^{}_1$ varying in the range from  1 to 10 TeV. It turns out that, for the parameter space consistent with the observed value of $Y^{}_{\rm B}$, the allowed values of ${\rm BR}(\mu \to e \gamma)$ are well below the current upper bound.

\section{Non-trivial flavour structure of $\mu^{}_{\rm s}$ assisted leptogenesis}

In this section, we study the scenario that the $\mu^{}_{\rm s}$ matrix also contains some non-trivial flavor structure parallel to that in Eq.~(\ref{1.9}). With the help of Eq.~(\ref{1.4}), utilizing the light neutrino mass matrix and the $M^{}_{\rm D}$ matrix, we can derive the $\mu^{}_{\rm s}$ matrix accordingly. This derivation allows us to proceed with the subsequent calculations of baryon asymmetry.

We will consider two distinct scenarios. In the first scenario, we simply assume that there exists no gap between the energy scale underlying the ISS model and the sterile neutrino mass scale so that the RGE effects can be safely neglected, in which case we just need to consider the contributions of the non-trivial flavor structure of $\mu_{\rm s}$ to the mass splittings among different PD sterile neutrino pairs.
By introducing non-trivial flavor structure for $\mu^{}_{\rm s}$, distinctions among the three $S^{}_I$ singlets arise, effectively breaking the mass degeneracies among the three PD sterile neutrino pairs.
Consequently, the $(\mathcal{B}_{\pm})_{\alpha I }$ terms no longer satisfy the relationship in Eq.~(\ref{2.19}).
Additionally, as in previous calculations, we need to diagonalize the sterile neutrino mass matrix to conveniently calculate the generated baryon asymmetry. However, in the present scenario, the non-trivial flavor structure of $\mu^{}_{\rm s}$ adds complexity to the diagonalization process and alters the structure of the unitary matrices involved. Specifically, the unitary matrices $V^{}_1$ and $V^{}_2$ in Eq.~(\ref{2.1}) will exhibit notable differences and cannot be simply related by the relation $V^{}_1 = {\rm i}V^{}_2$ (which holds in the case that $\mu^{}_s$ takes a simple form as shown in Eq.~(\ref{1.8})). Subsequently, the Yukawa coupling matrices are no longer constrained by the relationships outlined in Eq.~(\ref{2.19}). As a result, the cancellations for the total CP asymmetry will be weakened, enhancing the potential to generate a sufficiently large baryon asymmetry.

\begin{figure*}
\centering
\includegraphics[width=4.8in]{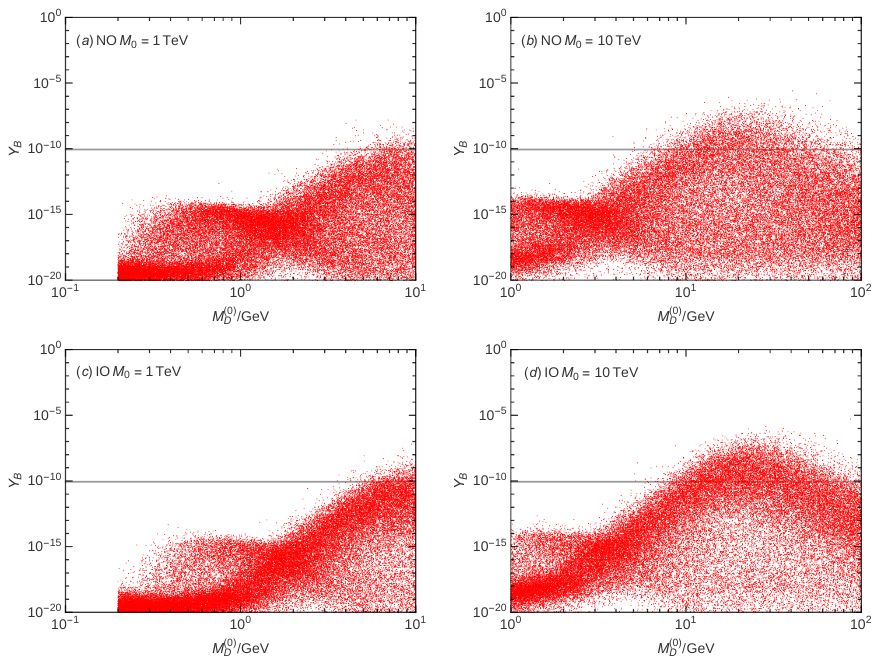}
\caption{ For the first scenario studied in section~4 where we only consider the contributions of the non-trivial flavor structure of $\mu_{\rm s}$ to the mass splittings among different PD sterile neutrino pairs, in the NO (IO) case, the allowed values of $Y^{}_{\rm B}$ as functions of $M^{(0)}_{\rm D}$ for some benchmark values of $M^{}_0$. The horizontal line stands for the observed value of $Y^{}_{\rm B}$. }
\label{fig3}
\end{figure*}

\begin{figure*}
\centering
\includegraphics[width=6.5in]{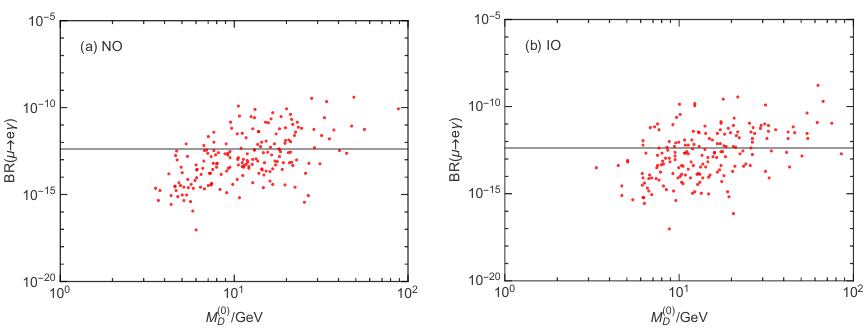}
\caption{ For the first scenario studied in section~4, in the parameter space that allows for a reproduction of the observed value of $Y^{}_{\rm B}$, the allowed values of ${\rm BR}(\mu \to e \gamma)$ as functions of $M^{(0)}_{\rm D}$  in the NO (a) and IO (b) cases. The horizontal line stands for the current upper bound on ${\rm BR}(\mu \to e \gamma)$. }
\label{fig4}
\end{figure*}

For the present scenario, Figure~\ref{fig3} has shown the allowed values of $Y^{}_{\rm B}$ as functions of $M^{(0)}_{\rm D}$ (the common overall scale of $M^{}_{\rm D}$) for the benchmark values of $M^{}_0 = 1$ and 10 TeV.
These results are obtained for the following parameter settings: for the neutrino mass squared differences and neutrino mixing angles, we employ the global-fit results for them \cite{global1, global2}. For the two Majorana CP phases of the neutrino mixing matrix, we allow them to vary in the range 0---$2\pi$, while the Dirac CP phase is constrained by the TM1 mixing scheme realized in the scenario considered in this paper. The lightest neutrino mass ($m^{}_1$ or $m^{}_3$ in the NO or IO case) is varied in the range between $0.001$ eV and $0.1$ eV.
To maintain the general form of $M^{}_{\rm D}$, we allow the absolute values of its matrix elements to fluctuate within a range of 0.5 to 2 times $M^{(0)}_{\rm D}$. To conform to the essential idea of the ISS model, we require $\mu^{}_{\rm s}$ to be at least two orders of magnitude smaller than $M^{}_{\rm D}$, and $M^{}_{\rm D}$ to be at least two orders of magnitude smaller than $M^{}_{\rm R}$.
The results show that in both the NO and IO cases the observed value of $Y^{}_{\rm B}$ has chance to be successfully reproduced.
It can be seen that the observed value of $Y^{}_{\rm B}$ has a relatively higher probability of being successfully reproduced for $M^{(0)}_{\rm D}$ around $1$ GeV.
Overall, it can be observed that as $M^{}_0$ varies in the range of 1 to 10 TeV, the observed value of $Y^{}_{\rm B}$ can be successfully reproduced for $M^{(0)}_{\rm D}$ in the range of $1$ to $10$ GeV.
For this scenario, we have also investigated the consequences of the model with respect to the $\mu \to e \gamma $ process. The results are shown in Figure~\ref{fig4}(a) and (b) (for the NO and IO cases, respectively). The results show that the parameter range of $M^{(0)}_{\rm D} \gtrsim 4$ GeV is  excluded by the constraint from the $\mu \to e \gamma $ process, leaving us with $2 \lesssim M^{(0)}_{\rm D}/{\rm GeV} \lesssim 4$. This stronger constraint from $\mu \to e \gamma $, compared to the case with a trivial $\mu_s$ structure, stems from the non-trivial flavor structure of $\mu_s$ which modifies the relationship between the diagonalization matrices $V_1$ and $V_2$. As a result, the condition $K_{\alpha 1} = {\rm i}K_{\alpha 4}$ no longer holds within each PD pair, thereby weakening the cancellation between their contributions to the cLFV amplitude. This leads to an enhanced branching ratio and consequently to a more stringent experimental constraint, bringing this scenario around the current experimental sensitivity and making it testable.

\begin{figure*}
\centering
\includegraphics[width=5in]{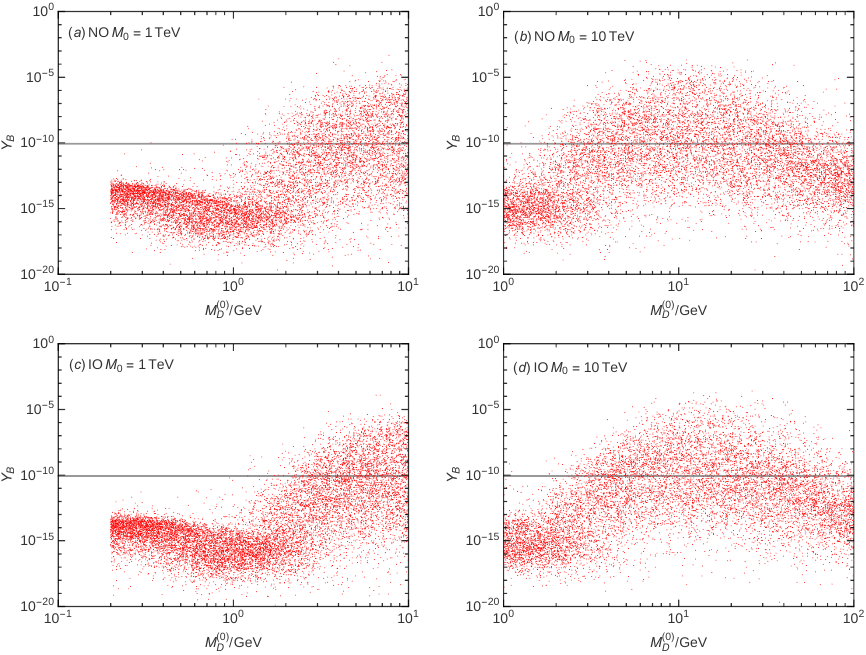}
\caption{ For the second scenario studied in section~4 where we take into account both the contributions of the RGE effects and $\mu^{}_{\rm s}$ to the mass splittings among different PD sterile neutrino pairs, in the NO (IO) case, the allowed values of $Y^{}_{\rm B}$ as functions of $M^{(0)}_{\rm D}$ for some benchmark values of $M^{}_0$. The horizontal line stands for the observed value of $Y^{}_{\rm B}$. }
\label{fig5}
\end{figure*}

\begin{figure*}
\centering
\includegraphics[width=6.5in]{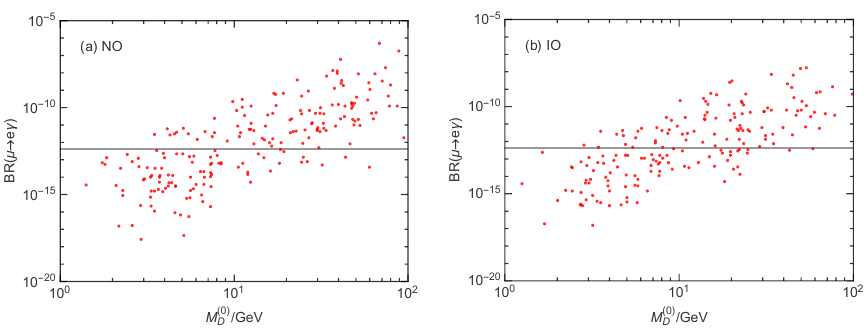}
\caption{ For the second scenario studied in section~4, in the parameter space that allows for a reproduction of the observed value of $Y^{}_{\rm B}$, the allowed values of ${\rm BR}(\mu \to e \gamma)$ as functions of $M^{(0)}_{\rm D}$  in the NO (a) and IO (b) cases. The horizontal line stands for the current upper bound on ${\rm BR}(\mu \to e \gamma)$. }
\label{fig6}
\end{figure*}

In the second scenario, we assume that there is a considerable gap between the energy scale underlying the ISS model and the sterile neutrino mass scale so that the RGE effects may become nonnegligible.
In this case, we take into account both the contributions of the RGE effects and $\mu^{}_{\rm s}$ to the mass splittings among different PD sterile neutrino pairs.
Although the contributions of the RGE effects are comparable to or exceed those of $\mu^{}_{\rm s}$, the latter provides an essential contribution. As noted previously, this structure complicates the diagonalization procedure. It is worth reiterating that this leads to notable differences between the unitary matrices $V^{}_1$ and $V^{}_2$ in Eq.~(\ref{2.1}), specifically invalidating the relation $V^{}_1 = {\rm i}V^{}_2$ that holds for the simple form of $\mu^{}_s$ in Eq.~(\ref{1.8}). Consequently, the Yukawa coupling matrices are no longer constrained by the relationships in Eq.~(\ref{2.19}). For this scenario, we have shown the allowed values of $Y^{}_{\rm B}$ as functions of $M^{(0)}_{\rm D}$ in Figure~\ref{fig5}(a) and (b) (for the NO and IO cases, respectively). It is found that the parameter space that allows for a reproduction of the observed value of $Y^{}_{\rm B}$ is broader than that shown in Figure~\ref{fig3}, where only the contributions of $\mu_{\rm s}$ were considered. Overall, it can be observed that as $M^{}_0$ varies in the range of 1 to 10 TeV, the observed value of $Y^{}_{\rm B}$ can be successfully reproduced for $M^{(0)}_{\rm D}$ in the range of $1$ to $10$ GeV. Figure~\ref{fig6}(a) and (b) (for the NO and IO cases, respectively) have shown the consequences of the model with respect to the $\mu \to e \gamma $ process. The results show that the parameter range of $M^{(0)}_{\rm D} \gtrsim 2$ GeV is  excluded by the constraint from the $\mu \to e \gamma $ process, leaving us with $1 \lesssim M^{(0)}_{\rm D}/{\rm GeV} \lesssim 2$.

\section{Summary}

The type-I seesaw mechanism provides an elegant explanation for both the neutrino masses and the baryon asymmetry of the Universe. However, the traditional type-I seesaw mechanism and associated leptogenesis mechanism have the drawback that the newly introduced particle states are too heavy to be directly accessed by foreseeable experiments. In this connection, the inverse seesaw (ISS) model provides an attractive framework that can naturally explain the smallness of neutrino masses while accommodating some sterile neutrinos potentially accessible at present or future experiments. However, in generic ISS models with hierarchical pseudo-Dirac (PD) sterile neutrino pairs (in which case the resonance effects for leptogenesis only take place within each pair of PD sterile neutrinos, in contrast with the scenario considered in this paper), the generation of the observed baryon asymmetry of the Universe via the leptogenesis mechanism is extremely challenging.

In this paper, we have investigated rescuing leptogenesis in the ISS model with the help of non-Abelian flavor symmetries which have the potential to explain the observed peculiar neutrino mixing pattern.
The basic idea is that the non-Abelian flavor symmetries can naturally enforce mass degeneracies among the three PD sterile neutrino pairs. After these degeneracies are broken in a proper way, resonant leptogenesis among different PD sterile neutrino pairs can naturally arise, potentially enhancing the generated baryon asymmetry. To be concrete, thanks to the non-Abelian flavor symmetries, $M^{}_{\rm R}$ and $\mu^{}_s$ can naturally take the forms in Eqs.~(\ref{1.7}, \ref{1.8}) which enforce mass degeneracies among the three PD sterile neutrino pairs, while  $M^{}_{\rm D}$ can naturally take a form in Eq.~(\ref{1.9}) which realizes the popular TM1 neutrino mixing. Then, we have considered two well-motivated approaches to generate the desired mass splittings: (i) they can be naturally induced by the renormalization group evolution (RGE) effects; (ii) when the mass matrix $\mu^{}_{\rm s}$ also contains some non-trivial flavor structure, the flavor non-universality of $\mu^{}_{\rm s}$ will break the mass degeneracies among the three PD sterile neutrino pairs.

In the first scenario, the RGE effects break the mass degeneracies of three RHNs, which finally result in mass splittings among the three PD sterile neutrino pairs. Figure 1(a) and (b) demonstrate that in both the NO and IO cases, the observed value of $Y^{}_{\rm B}$ can be successfully reproduced.
For smaller values of $M^{}_0$ (e.g., $M^{}_0=1$ TeV), relatively small values of $\mu^{}_0$ (around $10^3$ eV) are needed to reproduce the observed value of $Y^{}_{\rm B}$. As $M^{}_0$ increases, the scale of $\mu^{}_0$ needed to reproduce the observed $Y^{}_{\rm B}$ also rises. For this scenario, we have also investigated the consequences of the model with respect to the $\mu \to e \gamma $ process. It turns out that, for the parameter space consistent with the observed value of $Y^{}_{\rm B}$, the allowed values of ${\rm BR}(\mu \to e \gamma)$ are well below the current upper bound.

In the second scenario, the non-trivial flavor structure of $\mu^{}_{\rm s}$ breaks the mass degeneracies of three $S^{}_I$ singlets, which finally result in mass splittings among the three PD sterile neutrino pairs. When we simply assume there exists no gap between the energy scale underlying the ISS model and the sterile neutrino mass scale, the RGE effects can be safely neglected. For this scenario, we have only considered the contributions $\mu_{\rm s}$ to the mass splittings among different PD sterile neutrino pairs. Figure~\ref{fig3} shows that the observed value of $Y_{\rm B}$ can be successfully reproduced in both the NO and IO cases for $M^{(0)}_{\rm D}$ in the range of $1$ to $10$ GeV. When a considerable gap between the energy scale underlying the ISS model and the sterile neutrino mass scale, the RGE effects may become nonnegligible. For this scenario, we have considered both the contributions of the RGE effects and $\mu_{\rm s}$. Figure~\ref{fig5} shows that the parameter space that allows for a reproduction of the observed value of $Y^{}_{\rm B}$ is broader than that shown in Figure~\ref{fig3}, where only the contributions of $\mu_{\rm s}$ were considered. For these two scenarios, it turns out that the constraint from the $\mu \to e \gamma $ process can help us exclude some parameter space that allows for a reproduction of the observed value of $Y^{}_{\rm B}$.

\vspace{0.5cm}

\underline{Acknowledgments} \vspace{0.2cm}

This work was supported in part by the National Natural Science Foundation of China under Grant No. 12475112, Liaoning Revitalization Talents Program under Grant No. XLYC2403152, and the Basic Research Business Fees for Universities in Liaoning Province under Grant No. LJ212410165050.

\vspace{0.5cm}

\underline{Data availability} \vspace{0.2cm}

The data are not publicly available. The data are available from the authors upon reasonable request.

\end{document}